\documentclass{appolb}
\usepackage{graphicx,amssymb}
\usepackage{xcolor}

\begin{document}
\title{Effective shear and bulk viscosities of the quark-gluon plasma: QCD versus heavy-ion data%
\thanks{Presented at Quark Matter 2022}%
}
  \author{Fernando G. Gardim 
\address{Instituto de Ci\^encia e Tecnologia, Universidade Federal de Alfenas, 37715-400 Po\c cos de Caldas, MG, Brazil}
\\[3mm]
{Jean-Yves Ollitrault
\address{Universit\'e Paris-Saclay, CNRS, CEA, Institut de physique th\'eorique, 91191, Gif-sur-Yvette, France}}
}
\maketitle
\begin{abstract}
  In recent years, there has been a significant effort to extract the tempe\-rature-dependent shear ($\eta/s$) and bulk ($\zeta/s$) viscosity over entropy ratios of the quark-gluon plasma from a global comparison of heavy-ion data with results of hydrodynamic simulations. 
  However, anisotropic flow, which is arguably the most sensitive probe of viscosity, is only sensitive to an {\it effective\/} viscosity over entropy ratio, which is obtained by taking a weighted average over the temperature, and summing the contributions of shear and bulk.
  We estimate this effective viscosity using existing first-principles calculations, which give $0.17<(\eta/s)_{\rm eff}<0.21$, and  $(\zeta/s)_{\rm eff}<0.08$, implying that the damping of anisotropic flow at the LHC is mostly due to shear viscosity. 
  The values extracted from global data analyses are compatible with these theory predictions. 
\end{abstract}
  
\section{Introduction}
The quark-gluon plasma produced in ultrarelativistic heavy-ion collisions at RHIC and LHC is the most strongly-interacting medium ever produced in the laboratory, and strong interactions typically imply a small value of the shear viscosity ($\eta$) to entropy density ($s$) ratio $\eta/s$~\cite{Policastro:2001yc}. 
Calculating the transport coefficients of QCD is notoriously difficult.
But it is hoped that they can be constrained using heavy-ion experimental data, through a comparison with results of hydrodynamic simulations, which use transport coefficients as input. 
One expects $\eta/s$ to depend on temperature, with a minimum near a critical point or a rapid crossover~\cite{Csernai:2006zz}, and such a temperature dependence is included in hydrodynamic calculations~\cite{Niemi:2012ry}. 
The potential importance of bulk viscosity, $\zeta$, has also been emphasized~\cite{Bozek:2009dw,Ryu:2015vwa}.
Unlike shear viscosity, it is expected to increase near a critical point \cite{Karsch:2007jc}.

It may seem natural to try and extract $\eta/s(T)$ and $\zeta/s(T)$ from a global comparison of experimental data with results of hydrodynamic simulations, and several attempts have been made in this direction~\cite{Bernhard:2019bmu,Nijs:2020roc,JETSCAPE:2020mzn}, but error bars are still rather large. 
An alternative approach is to consider observables one by one and characterize their dependence on transport coefficients.
It is well known that this dependence is strongest for anisotropic flow coefficients, in particular elliptic flow, $v_2$, and triangular flow, $v_3$. 
We have recently shown through detailed hydrodynamic simulations~\cite{Gardim:2020mmy} that each of these Fourier harmonics depends on $\eta/s(T)$ and $\zeta/s(T)$ only through a single quantity, which we refer to as an effective viscosity.
The effective viscosity only depends on collision energy, not on system size or collision centrality. 
Its definition is recalled in Sec.~\ref{s:effective}. 
Thus, the sensitivity of anisotropic flow at the LHC to transport coefficients is encapsulated in two effective viscosities, one for $v_2$, one for $v_3$.
In Sec.~\ref{s:eval}, we evaluate these effective viscosities from first-principles QCD calculations, and from Bayesian analyses of heavy-ion data.

\section{Viscous damping of $v_n$ in relativistic hydrodynamics}
\label{s:effective}

Relativistic hydrodynamics describes the evolution of an interacting system over large space-time scales.
It is usually formulated as an expansion~\cite{Baier:2007ix} in $\lambda/R$, where $\lambda$ is the mean free path or an equivalent microscopic scale, and $R$ is the large scale, here the nuclear radius. 
Ideal hydrodynamics is the leading term in the expansion, while shear and bulk viscosities are the coefficients of the first-order correction, of relative order $\lambda/R$. 
The validity of hydrodynamics itself implies that the first-order correction to an arbitrary observable ${\cal O}$ is small. 
We evaluate its magnitude in the following way.
We solve ideal and viscous hydrodynamics with the exact same initial conditions, up to the overall normalization of the entropy density, which we fix in such a way that the final particle multiplicity is identical for both calculations. 
The dependence of ${\cal O}$ on $\eta$ and $\zeta$ can be expanded to first order, and the most general expression is 
\begin{equation}
  \label{linearity}
  {\cal O}({\rm viscous})=  {\cal O}({\rm ideal})\left(
  1+\int w_{\cal O}^{(\eta)}(T)\frac{\eta}{s}(T)dT
  +\int w_{\cal O}^{(\zeta)}(T)\frac{\zeta}{s}(T)dT
  \right),
\end{equation}
where the weights $w_{\cal O}^{(\eta)}(T)$ and $w_{\cal O}^{(\zeta)}(T)$ quantify the sensitivity of the observable ${\cal O}$  to shear and bulk viscosities at temperature $T$.
The integral runs over the range of temperatures spanned by the hydrodynamic calculation.
The lower bound is the freeze-out temperature $T_f$ at which the fluid is converted into particles. 
The evolution of the fluid for $T>T_f$ is fully determined by the relativistic Navier-Stokes equations.
On the other hand, how the fluid is converted into particles at freeze-out depends on the details of microscopic interactions~\cite{Dusling:2011fd}. 
We write the weight $w_{\cal O}^{\eta,\zeta}(T)$ as the sum of a smooth function for $T>T_f$, and a discrete contribution proportional to $\delta(T-T_f)$, corresponding to the viscous correction at freeze-out.
We expect that most, if not all, the model dependence of our calculation lies in the discrete part.

\begin{figure}[htb]
\centerline{%
\includegraphics[width=12.5cm]{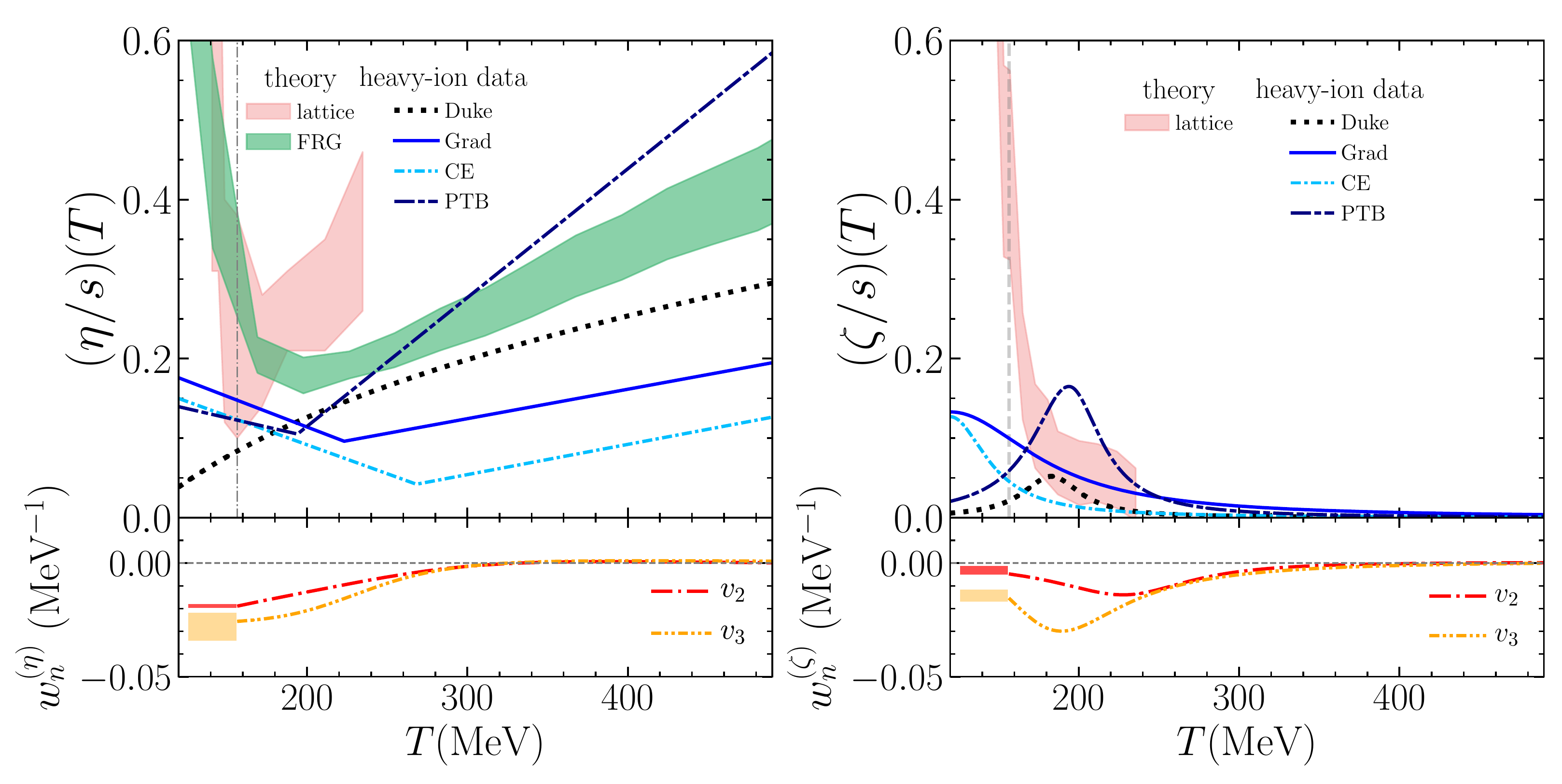}}
\caption{Top panel: Shear (left) and bulk (right) viscosity over entropy ratios versus temperature. Bands correspond to lattice QCD ~\cite{Astrakhantsev:2017nrs,Astrakhantsev:2018oue} or FRG calculations~\cite{Christiansen:2014ypa}, where we set $T_c=156$~MeV~\cite{HotQCD:2018pds}. 
  Lines correspond to global analyses of heavy-ion data by the Duke group~\cite{Bernhard:2019bmu} and the JETSCAPE Collaboration~\cite{JETSCAPE:2020mzn}.
  The latter includes three different prescriptions for the emission of particles at freeze-out, referred to as  Grad, Chapman-Enskog (CE) and Pratt-Torrieri-Bernhard (PTB).
  Bottom panel: Values of the weights entering Eq.~(\ref{linearity}) for $v_2$ and $v_3$.
  The shaded boxes represent the magnitude of the discrete part $\propto\delta(T-T_f)$, corresponding to the viscous correction at freeze-out. 
}
\label{Fig:tdependence}
\end{figure}

We apply this approach to anisotropic flow.
The lower panels of Fig.~\ref{Fig:tdependence} display the weights corresponding to elliptic flow,
$v_2$, and triangular flow, $v_3$, averaged over $p_t$ and over the pseudorapidity range $|\eta|<0.5$, for Pb+Pb collisions at $\sqrt{s_{\rm NN}}=5.02$~TeV in the 0-5\% centrality window~\cite{Gardim:2020mmy}. 
They are negative for most temperatures, implying a viscous suppression of $v_n$.
The shaded boxes represent the viscous correction at freeze-out.
Their area is much smaller than the area under the curve, which implies that the determination of the viscous suppression is robust.
[Note that this is no longer the case at RHIC energies~\cite{Gardim:2020mmy}.] 
The robustness is due to the fact that we integrate $v_n$ over transverse momentum.
Differential observables~\cite{Nijs:2020roc} are by construction more sensitive to the freeze-out procedure. 

We define the effective shear viscosity as the average over the temperature with the corresponding weight:
\begin{equation}
  \label{defeffective}
  \left(\frac{\eta}{s}\right)_{n,\rm eff}=\frac{\int_{T_f}^\infty (\eta/s)(T)w^{(\eta)}_n(T)dT}{\int_{T_f}^\infty w^{(\eta)}_n(T)dT},
\end{equation}
and the effective bulk viscosity $(\zeta/s)_{n,\rm eff}$ is defined by the same equation where one replaces $w^{(\eta)}_n(T)$  with $w^{(\zeta)}_n(T)$. 
Note that there is one effective viscosity for each observable, hence the subscript $n$. 

With these definitions, Eq.~(\ref{linearity}) gives, for $v_2$ and $v_3$,
\begin{eqnarray}
  \label{v2v3}
  v_2({\rm viscous})&=&v_2({\rm ideal})\left(1-1.34\left(\frac{\eta}{s}\right)_{\rm 2,eff}-1.30\left(\frac{\zeta}{s}\right)_{\rm 2,eff}\right)\cr
  v_3({\rm viscous})&=&v_3({\rm ideal})\left(1-2.33\left(\frac{\eta}{s}\right)_{\rm 3,eff}-2.61\left(\frac{\zeta}{s}\right)_{\rm 3,eff}\right)
\end{eqnarray}
where the numerical coefficients in front of the effective viscosities are the denominators in Eq.~(\ref{defeffective}). 
Note that the viscous suppression is larger by a factor $\sim 2$ for $v_3$ than for $v_2$. 
The coefficients in Eq.~(\ref{v2v3}) are almost identical for shear and bulk, which means that $v_n$ only depends on the total effective viscosity, defined as the sum of shear and bulk effective viscosities. 

\section{Evaluating effective viscosities}
\label{s:eval}

\begin{figure}[htb]
\centerline{%
\includegraphics[width=6cm]{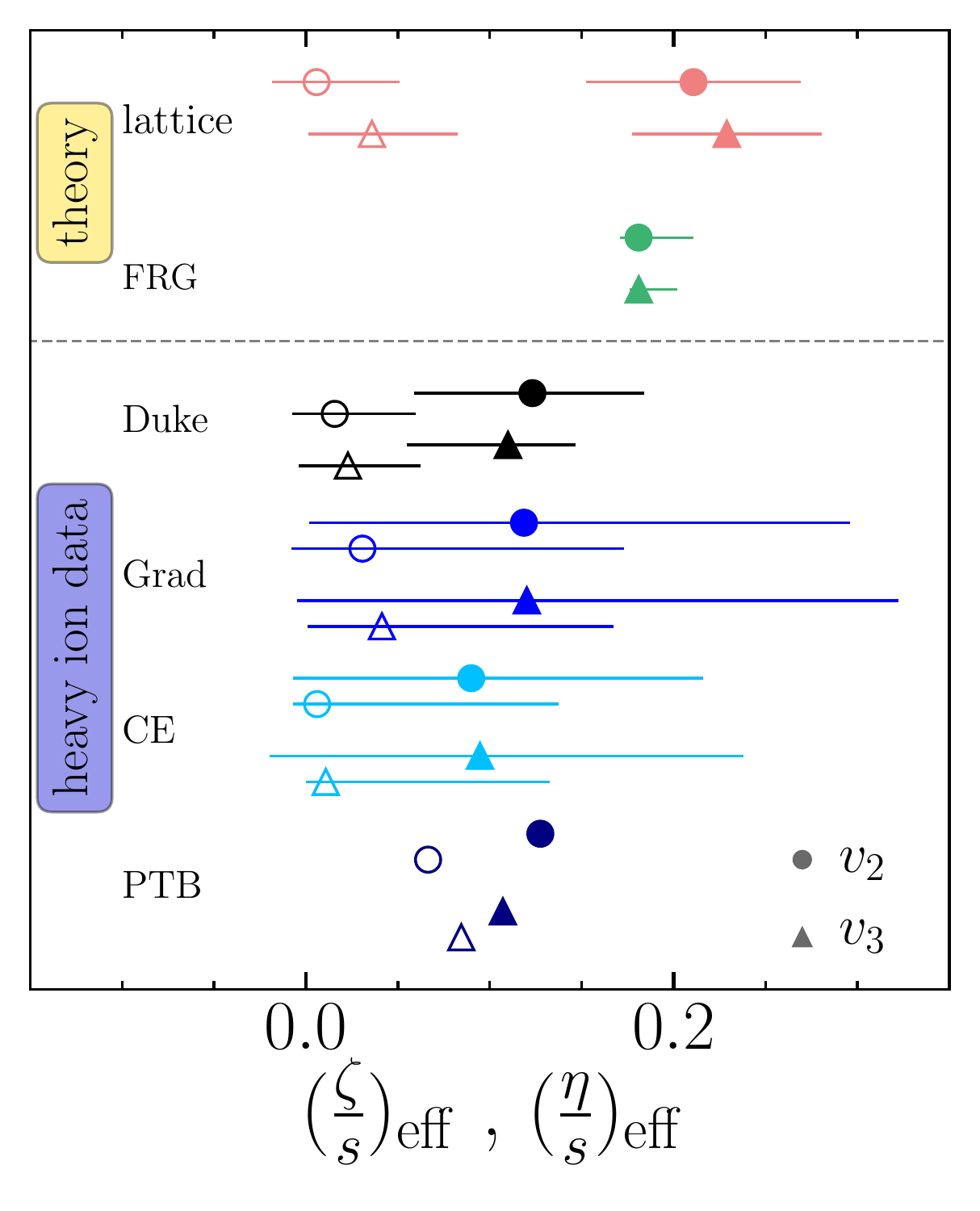}}
\caption{Effective bulk (open symbols) and shear (closed symbols)  viscosities, obtained by multiplying the temperature-dependent viscosities (upper panel of Fig.~\ref{Fig:tdependence}) with the weights (lower panel of Fig.~\ref{Fig:tdependence}) and integrating over the temperature. 
}
\label{Fig:effective}
\end{figure}

The shear and bulk viscosities of QCD have been calculated with reasonable accuracy at low temperature~\cite{Noronha-Hostler:2008kkf} and at high temperature~\cite{Arnold:2003zc,Ghiglieri:2018dib}. 
In order to evaluate the effective viscosities, however, we need to integrate over all temperatures, the important range being $150\lesssim T\lesssim 350$~MeV where the weights $w_n^{(\eta,\zeta)}(T)$ are largest.
Few calculations cover this whole range. 
For the shear viscosity, we use results using Functional Renormalization Group (FRG) methods~\cite{Christiansen:2014ypa}, where quarks are included, and results from lattice QCD~\cite{Astrakhantsev:2017nrs}, where quarks are not included but their effect is estimated.  
For bulk viscosity, we use lattice QCD calculations~\cite{Astrakhantsev:2018oue} where quarks are not included. 
These calculations are displayed as shaded bands in the upper panels of Fig.~\ref{Fig:tdependence}.
The lines in this figure display viscosities inferred from global analyses of heavy-ion data by the Duke group~\cite{Bernhard:2019bmu} and the JETSCAPE Collaboration~\cite{JETSCAPE:2020mzn}.
We only show the central values, not the error bands, but we use the error bands to evaluate the errors on effective viscosities (see Fig.~\ref{Fig:effective}). 
The differences between the lines labeled Grad, CE, PTB illustrate the sensitivity of hydrodynamic calculations to theoretical uncertainties in modeling the freeze-out stage.

We then evaluate the effective viscosities defined by Eq.~(\ref{defeffective}), which are displayed in Fig.~\ref{Fig:effective}. 
They are almost identical for $v_2$ and for $v_3$, which reflects the fact that the weights defining these viscosities are very similar for both observables, up to an overall normalization (see lower panel of Fig.~\ref{Fig:tdependence}).
The effective viscosities inferred from the Bayesian analysis of experimental data by the JETSCAPE collaboration depend little on the ansatz used for the particlization of the fluid (Grad, CE, or PTB).
This illustrates our point that effective viscosities encapsulate the information that can be inferred from $v_n$ data. 
Results from the earlier Duke analysis seem to have smaller errors, but they are likely to be underestimated~\cite{JETSCAPE:2020mzn}. 
One sees that Bayesian analyses give in general $(\zeta/s)_{\rm eff}<(\eta/s)_{\rm eff}$, although with large error bars. 
Note that $v_2$ and $v_3$ only depend on the sum of bulk and shear effective viscosities, hence they do not allow to disentangle shear from bulk. 
It would be useful to extend our approach to other observables, such as the mean transverse momentum $\langle p_t\rangle$, for which the effects of shear and bulk viscosity go in opposite directions~\cite{Gardim:2019xjs}

The first-principles calculation of~\cite{Christiansen:2014ypa} gives a narrow range for the effective shear viscosity, $0.17<(\eta/s)_{\rm eff}<0.21$, which is compatible with the lattice calculation, and with the JETSCAPE result within error bars. 
It is interesting to note that the lattice calculation also predicts $(\zeta/s)_{\rm eff}<(\eta/s)_{\rm eff}$, in line with general expectations that bulk viscosity should be negligible compared to shear viscosity~\cite{Arnold:2006fz}. 

\textbf{Acknowledgements:} FGG was supported by CNPq grant 306762/2021-8, by INCT-FNA grant 464898/2014-5 and FAPESP grant 2018/24720-6.

\end{document}